\newcommand\beq{\begin{eqnarray}}
\newcommand\eeq{\end{eqnarray}}
\def\lsim{\mathrel{\rlap{\lower4pt\hbox{$\sim$}}
    \raise1pt\hbox{$<$}}}                
\def\gsim{\mathrel{\rlap{\lower4pt\hbox{$\sim$}}
    \raise1pt\hbox{$>$}}}

\documentclass[
amsmath,
prd,nofootinbib,floatfix,11pt,
]{revtex4}

\allowdisplaybreaks
\usepackage{graphicx}
\usepackage{setspace}

\begin{document}

\title{\Large \baselineskip=20pt 
Shift in the LHC Higgs diphoton mass peak from\\ interference with background\\}

\author{Stephen P.~Martin}
\affiliation{
{\it Department of Physics, Northern Illinois University, DeKalb 
IL 60115}  and \\
{\it Fermi National Accelerator Laboratory, P.O. Box 500, Batavia 
IL 60510}
}

\begin{abstract}\normalsize \baselineskip=14pt 
The Higgs diphoton amplitude from gluon fusion at the LHC interferes with the
continuum background induced by quark loops. I investigate the effect of this interference on the position of the diphoton invariant mass peak used to help determine the Higgs mass. At leading order, the interference shifts the
peak towards lower mass by an amount of order 150 MeV or more, with the precise value dependent on the methods used to analyze and fit the data.
\end{abstract}


\maketitle


\baselineskip=15.4pt

\setcounter{footnote}{1}
\setcounter{figure}{0}
\setcounter{table}{0}

The ATLAS and CMS collaborations at the CERN Large Hadron Collider (LHC)
have recently announced the discovery of a resonance with production rates 
and decay branching ratios that are at least approximately consistent with the 
Standard Model Higgs scalar boson \cite{ATLASHiggs,CMSHiggs}. In this paper, 
this resonance will be assumed to be indeed the Standard Model Higgs $H$. The  
detailed properties of $H$, including measurements of its spin and CP 
quantum numbers, mass, production cross-sections in various channels, and
branching ratios will be the focus of long-term experimental and theoretical 
investigations. The mass of $H$ is currently estimated to be 
$125.3 \pm 0.4 \mbox{(stat)} \pm 0.5 \mbox{(syst)}$ GeV by CMS and 
$126.0 \pm 0.4 \mbox{(stat)} \pm 0.4 \mbox{(syst)}$ by ATLAS.
After the accumulation of much more data, the experimental uncertainty in the 
mass may be reduced to perhaps \cite{ATLASTDR} 0.1 GeV, motivating efforts to reduce theoretical sources of error as much as is possible.

The purpose of this note is to point out the effect of signal-background 
interference on the determination of the Higgs mass from data in the diphoton
final state. 
The largest production cross-section for $H$ is from gluon-gluon
fusion $gg \rightarrow H$ \cite{Georgi:1977gs}, through loop diagrams mediated 
by quarks, with the top quark providing by far the biggest contribution. 
A tremendous effort has been expended in computing higher order corrections,
including next-to-next-to-leading order in QCD 
\cite{Dawson:1990zj,Djouadi:1991tka,Spira:1995rr,Harlander:2002wh,Anastasiou:2002yz,Ravindran:2003um,Anastasiou:2005qj}, 
next-to-leading order in 
electroweak couplings \cite{Aglietti:2004nj,Actis:2008ug,Anastasiou:2008tj}, 
and next-to-next-to-leading logs in soft gluon resummation
\cite{Catani:2003zt,deFlorian:2011xf}; for reviews see 
\cite{deFlorian:2009hc,Dittmaier:2011ti,Dittmaier:2012vm,Anastasiou:2012hx}. 
The 
rare but clean decay $H\rightarrow \gamma\gamma$ 
\cite{Ellis:1975ap,Shifman:1979eb,Gunion:1985dj,Ellis:1987xu,Gunion:1987ke,Djouadi:1997yw}
is also mediated by loop diagrams. The excellent electromagnetic energy resolution of the ATLAS and CMS detectors makes this channel, along with $H \rightarrow ZZ^{(*)} \rightarrow \ell^+\ell^-\ell^{\prime +}\ell^{\prime -}$,
one of the two best ways to determine $M_H$. The 
largest contribution to the $H \rightarrow \gamma\gamma$ amplitude comes 
from the $W$ loop, with a subdominant contribution of the opposite sign coming from the top quark.
(In 
this paper, the loop effects of $t,b,c$ quarks and the $\tau$ lepton, including 
their mass dependences, are included in the $H$ production and decay amplitudes.) 
The complete process $gg \rightarrow H \rightarrow \gamma\gamma$ is therefore 
of 2-loop order.  It can interfere with the continuum background process
$gg \rightarrow \gamma\gamma$, which is mediated by quarks beginning at one-loop 
order. 

Dicus and Willenbrock found \cite{Dicus:1987fk} that 
the effect of the interference on the total $\gamma\gamma$ rate is very small
at leading order because the interference involving the real parts of the
amplitudes is odd in $\hat s$ 
(the invariant squared mass of the parton-level process) 
around $M_H$, while the imaginary part of the continuum 
$gg \rightarrow \gamma\gamma$
amplitude has a quark mass suppression for the helicity combinations that can interfere with Higgs exchange.  
Dixon and Siu have
shown \cite{Dixon:2003yb} that the most important interference effect on the cross-section
instead comes from the imaginary part of the continuum
amplitude $gg \rightarrow \gamma\gamma$ 
at 2-loops \cite{Bern:2001df} (which, for the 
$++\rightarrow ++$ and $--\rightarrow --$ polarization configurations, does not have the mass 
suppression for the complex phase found at 1-loop order),
and that it is destructive and typically of order 2-5\% depending on the scattering angle. 
In the present paper, I 
consider the orthogonal issue of the shift in the position of the diphoton 
peak invariant mass
distribution. I will show that the leading-order effect 
of the interference results in 
a downward shift of the $M_{\gamma\gamma}$ peak, of order 150 MeV or more, 
compared to the result one would obtain when interference is ignored. 
The precise magnitude of this shift will depend on the method used to analyze 
and fit the data.
Other studies of the effects of the interference of the 
Higgs with backgrounds include 
\cite{Binoth:2006mf,Campbell:2011cu,Kauer:2012hd} for $gg \rightarrow H \rightarrow W^+ W^-$,
\cite{Glover:1988fe,Passarino:2012ri,Kauer:2012hd} for $gg \rightarrow H \rightarrow ZZ$, and
\cite{Dixon:2008xc} for $\gamma\gamma \rightarrow H \rightarrow b \overline b$
at a photon collider.

In making a precise determination of the Higgs mass, one must first choose a prescription to define it.
Consider the renormalized propagator for $H$,
\beq
\frac{i}{\hat s - m_H^2 - \Pi_H(\hat s)} 
= \frac{i F_H(\hat s)}{\hat s - M_H^2 + i M_H \Gamma_H} ,
\eeq
where $m_H$ is the tree-level mass and $\Pi_H$ is the 1PI self-energy function,
and $F_H(\hat s)$ is a function that is slowly varying and satisfies $F_H 
\approx 1$ in the resonance region. The complex pole mass 
$M_H^2 - i M_H \Gamma_H$ is a gauge-invariant physical observable, with 
$\Gamma_H$ the width of the Higgs, and will be 
used in the following to define the mass $M_H$. In the following, I will
ignore the variation in $F_H$ from 1 for simplicity; it would have only a 
very small effect on the considerations below for $M_H \sim 125$ GeV.

The leading order matrix element for $gg\rightarrow \gamma\gamma$ 
including both non-resonant and Higgs resonant amplitudes can be written as 
\beq
{\cal M} = 
-\delta^{ab} \delta_{\lambda_1 \lambda_2} \delta_{\lambda_3\lambda_4}
\frac{A_{ggH} A_{\gamma\gamma H}}{\hat s - M_H^2 + i M_H \Gamma_H} 
+
\delta^{ab} 4 \alpha \alpha_S \sum_{q = u,d,s,c,b,t} e_q^2 
M^q_{\lambda_1\lambda_2\lambda_3\lambda_4} ,
\eeq
where $a,b = 1,\ldots,8$ are $SU(3)_c$ adjoint representation indices 
for the gluons, and
the circular polarizations labels $\pm$ are
$\lambda_1, \lambda_2$ for the incoming gluons and  $\lambda_3, \lambda_4$
for the outgoing photons. 
The 1-loop amplitudes for $H$ coupling to gluons and to
photons are
\beq
A_{ggH} &=& -\frac{\alpha_S}{8\sqrt{2} \pi v} \hat s
\sum_{q = t,b,c} F_{1/2} (4 m_q^2/\hat s),
\\
A_{\gamma\gamma H} &=& -\frac{\alpha}{4\sqrt{2} \pi v} \hat s \biggl [
F_1(4 m_W^2/\hat s) + 
\sum_{f = t,b,c,\tau} N_{c}^f e_f^2 F_{1/2} (4 m_f^2/\hat s) \biggr ]
,
\eeq
where $v = 174$ GeV is the Higgs vacuum expectation value, and $N_c^f=3$ (1)
for $f=$ quarks (leptons) with electric charge $e_f$ and mass $m_f$, and
\beq
F_1(x) &=& 2 + 3 x [1 + (2-x) f(x)],
\\
F_{1/2} (x) &=& -2 x [1 + (1-x) f(x)],
\\
f(x) &=& \Biggl \{ \begin{array}{ll}
[\arcsin (\sqrt{1/x})]^2, & x\geq 1 \quad \mbox{(for $t,W$)},
\\
-\frac{1}{4} \left [ \ln \left (\frac{1 + \sqrt{1-x}}{1 - \sqrt{1-x}}\right ) 
- i \pi \right ]^2,\phantom{xxx}
& x\leq 1 \quad \mbox{(for $b,c,\tau$)}.
\end{array}
\eeq
The 1-loop
matrix elements
$M^q_{\lambda_1\lambda_2\lambda_3\lambda_4}$ mediated by quarks 
$q$
are the same as found in $\gamma\gamma \rightarrow \gamma\gamma$ 
scattering \cite{Karplus:1950zz},
and are used here in the normalization and sign convention such that,
when $m_q^2 \ll \hat s$, the polarization configurations 
that can give a non-zero interference
with the Higgs-mediated amplitudes are:
\beq
M^q_{++--} = M^q_{--++} &=& 1,
\label{eq:Mppmm} \\
M^q_{++++} = M^q_{----} &=& -1 +
z \ln\left (\frac{1+z}{1-z}\right ) - \frac{1 + z^2}{4} \biggl
[\ln^2\left (\frac{1+z}{1-z} \right ) + \pi^2 \biggr ],
\label{eq:Mpppp}
\eeq
where $z = \cos\theta_{\rm CM}$, with $\theta_{\rm CM}$ 
the scattering angle in the diphoton center-of-momentum (CM) frame. 
Note that in this light quark limit, these amplitudes are real, while the 
polarizations that have non-trivial complex phases at 1-loop do 
not interfere with the $H$-mediated amplitude.
In the following the $u,d,s$ quarks are treated as massless and the 
full mass dependence of $t,b,c$ quarks is included, using 
the formulas in \cite{Costantini:1971cj,Combridge:1980sx}. 

The contributions to the LHC diphoton production cross-section at 
leading order, in excess of the pure 
continuum background, can then be written as
\beq
\frac{d^2\sigma_{pp \rightarrow \gamma\gamma}}{d(\sqrt{\hat s})\, dz}
= \frac{G(\hat s)}{128\pi \sqrt{\hat s} D(\hat s)} 
(N_H + N_{\rm int, Re} + N_{\rm int, Im})
\label{eq:sigmahint}
\eeq
where
\beq
N_H &=& |A_{ggH} A_{\gamma\gamma H}|^2,
\label{eq:NH}
\\
N_{\rm int, Re} &=&  
-(\hat s - M_H) 2 {\rm Re} [ A_{ggH} A_{\gamma\gamma H} A_{gg\gamma\gamma}^* ],
\label{eq:NintR}
\\
N_{\rm int, Im} &=&  
- M_H \Gamma_H 2 {\rm Im} [ A_{ggH} A_{\gamma\gamma H} A_{gg\gamma\gamma}^* ],
\label{eq:NintI}
\eeq
for the Higgs and real and imaginary interference contributions. Here
\beq
A_{gg\gamma\gamma} = 2 \alpha_S \alpha \sum_q e_q^2 (M^q_{++++} + M^q_{++--}),
\label{eq:Aggaa}
\eeq
and 
\beq
G(\hat s) = \int_{\hat s/s}^1 \frac{dx}{sx} g(x) g(\hat s/sx)
\eeq
is the gluon-gluon luminosity function, and
\beq
D(\hat s) = (\hat s - M_H^2)^2 + M_H^2 \Gamma^2_H .
\eeq
The numerical results below use $M_H = 125$ GeV and $\Gamma_H = 4.2$ MeV 
for purposes of presentation, even though the current experimental indications 
are for a slightly heavier $H$. The running $\overline{\rm MS}$ fermion masses 
at $Q=M_H$ are taken to be $m_t = 168.2$ GeV, $m_b = 2.78$ GeV, 
$m_c = 0.72$ GeV, $m_{\tau} = 1.744$ GeV, and
$\alpha = 1/127.5$. The gluon distribution function $g(x)$ and strong coupling 
$\alpha_S(Q)$
are taken from the MSTW2008 NLO set \cite{Martin:2009iq}, with $Q^2 = \hat s$.
Because the focus here is on the shift in the diphoton mass peak, the very 
small imaginary
interference term in eq.~(\ref{eq:NintI}) and its 2-loop counterpart
discussed in \cite{Dixon:2003yb} will be neglected here, 
since they are small and 
affect the overall size but not the shape of the invariant mass distribution.
Numerical results will be shown for the 2012 run energy $\sqrt{s} = 8$ TeV,
but the results on the shape (as opposed to the size) of the $M_{\gamma\gamma}$ 
distribution turn out to be nearly independent of the LHC beam
energy at leading order. 
This is because the $\sqrt{s}$ dependence enters 
only through $G(\hat s)$, which appears in 
front of both $N_H$ and $N_{\rm int, Re}$ in eq.~(\ref{eq:sigmahint}).

In the leading order calculation given here, changes in the choices of 
renormalization and factorization scale only affect the size of the 
diphoton distribution but not its shape. This is again because the 
factorization scale at leading order enters only through $G(\hat s)$ which 
is common to both $N_H$ and $N_{\rm int, Re}$, and also because both the 
continuum and resonance amplitudes are proportional to $\alpha \alpha_S$. 
However, one might expect a potentially large change from including 
genuine next-to-leading order effects (beyond the scope of the present 
paper), since the $K$-factors for both continuum and resonance 
diphoton production are known to be large.

\begin{figure}[!t]
\begin{minipage}[]{0.41\linewidth}
\begin{center}
\includegraphics[width=8.2cm,angle=0]{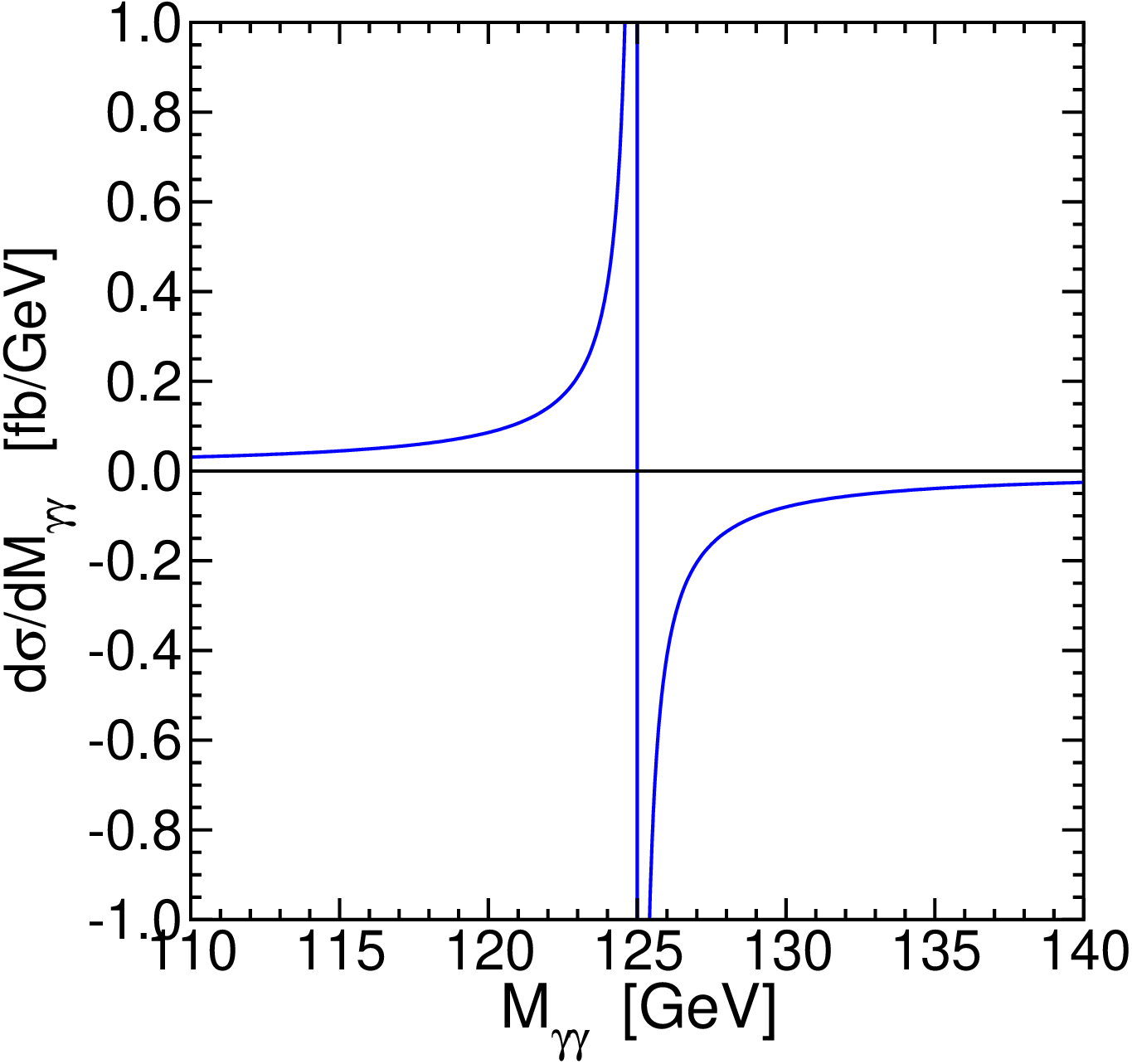}
\end{center}
\end{minipage}
\begin{minipage}[]{0.58\linewidth}
\begin{flushright}
\includegraphics[width=8.2cm,angle=0]{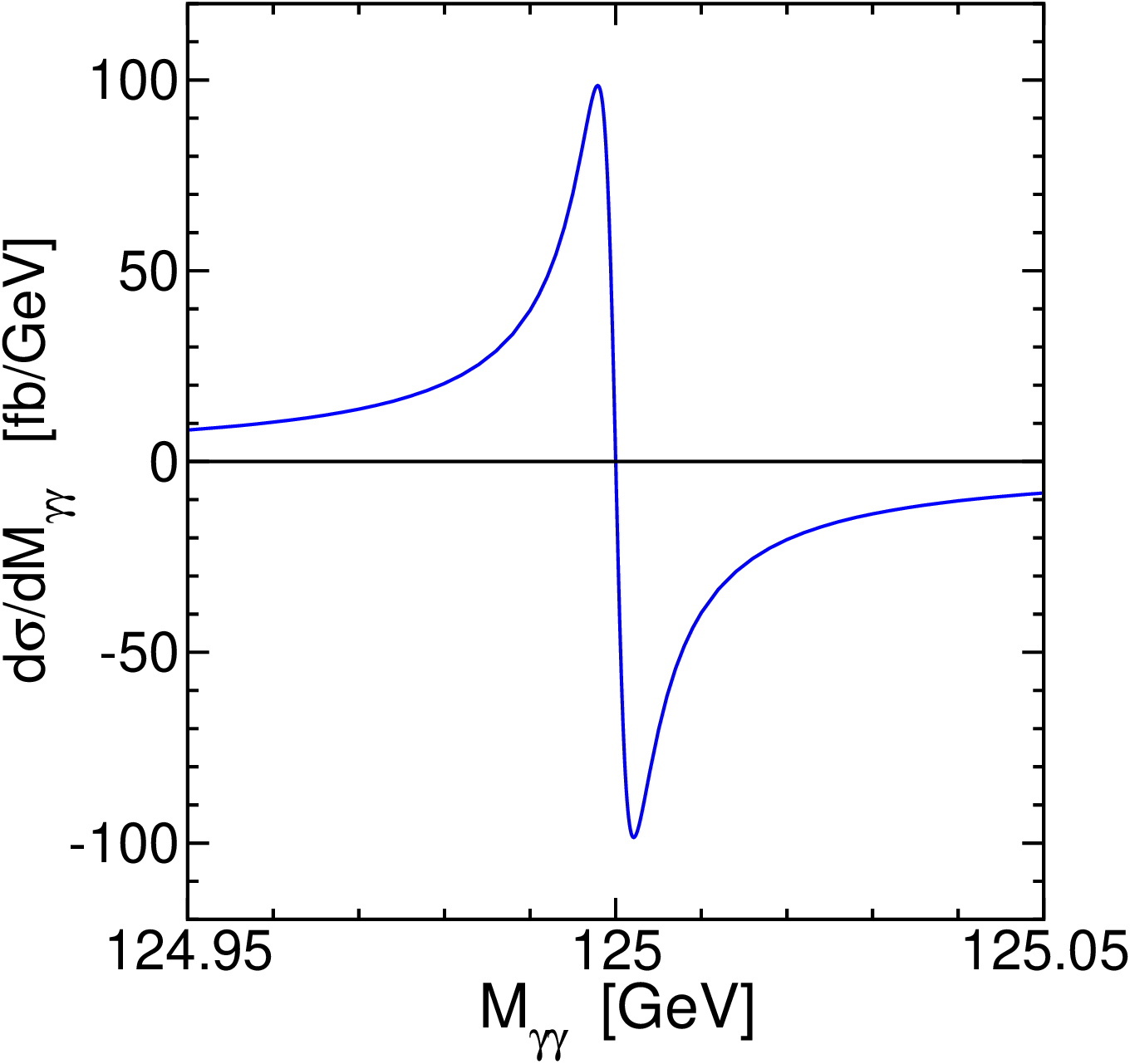}
\end{flushright}
\end{minipage}
\caption{\label{fig:unsmeared} The distribution of diphoton invariant masses
from the real interference
term in eq.~(\ref{eq:NintR}), as a function of 
$M_{\gamma\gamma} = \sqrt{\hat s}$, from eq.~(\ref{eq:sigmahint}),
before including experimental resolution effects.
The right panel is a close-up of the left panel, showing the 
maximum and minimum near $M_{\gamma\gamma} = M_H \pm \Gamma_H/2$.} 
\end{figure}

The  
factor of $\hat s - M_H^2$ in $N_{\rm int, Re}$  is odd about the Higgs peak, 
making its contribution to the total cross-section negligible 
when $\hat s$ is integrated over 
\cite{Dicus:1987fk,Dixon:2003yb}. However, the same
factor implies a slight excess for $M_{\gamma\gamma} = \sqrt{\hat s}$ 
below $M_H$ and a slight deficit above,
therefore pushing the peak to lower $M_{\gamma\gamma}$ than it would be if
interference were absent. 
This is shown 
first in the case without any experimental 
resolution effects for the photons, in Figure \ref{fig:unsmeared}.
The distribution shown is obtained from the real interference term in
eq.~(\ref{eq:NintR}), plugged in to
eq.~(\ref{eq:sigmahint}),
after integrating over $-1<z<1$ and dividing by 2 for identical photons.
The distribution shows a sharp peak and dip near $M_{\gamma\gamma} = 
M_H - \Gamma_H/2$ and $M_H + \Gamma_H/2$ respectively, but there are also long
tails due to the Breit-Wigner shape. [Using a different prescription for the width in the Breit-Wigner lineshape, such as the running-width prescription
with $D(\hat s) = (\hat s - M_H^2)^2 + \hat s [\Gamma_H(\hat s)]^2$, does not
significantly affect the results, because for a light Higgs boson the width term 
is only important very close to the resonance peak where the width term is 
nearly constant.]

\begin{figure}[!t]
\begin{minipage}[]{0.59\linewidth}
\begin{flushleft}
\includegraphics[width=9.0cm,angle=0]{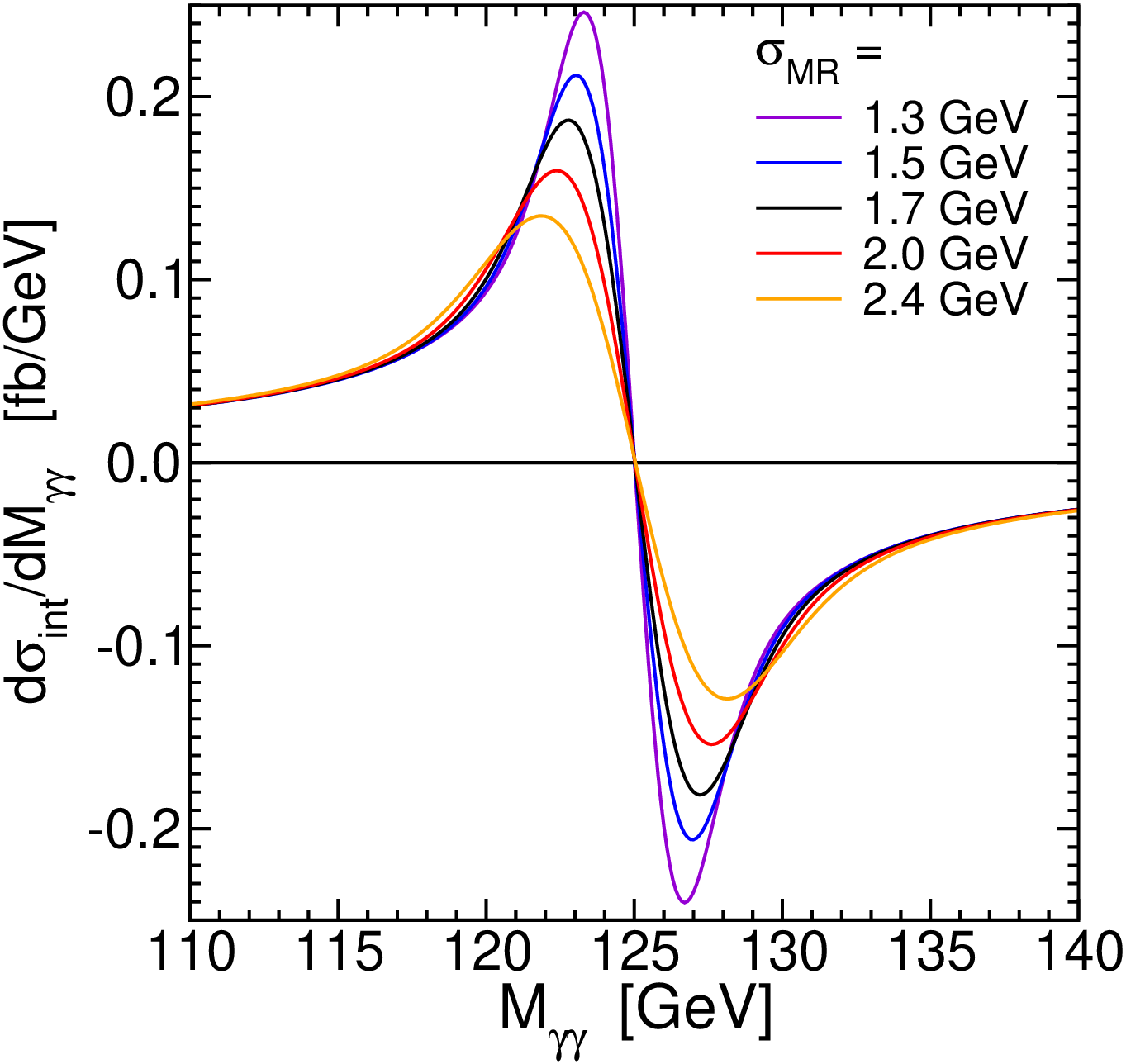}
\end{flushleft}
\end{minipage}
\begin{minipage}[]{0.40\linewidth}
\caption{\label{fig:diffsmeared} The distribution of diphoton invariant masses
from the real interference, as in Figure \ref{fig:unsmeared}, but now
smeared by various Gaussian mass resolutions with widths 
$\sigma_{\rm MR}$.}
\end{minipage}
\end{figure}

At the LHC, the photon energies are smeared by detector effects, in ways
that differ between the two experiments. A detailed treatment of these effects
is beyond the scope of this paper,
but as an approximation, 
Figure \ref{fig:diffsmeared} shows the same interference as in Figure 
\ref{fig:unsmeared},
but now convoluted with some representative Gaussian\footnote{In the real 
experiments, the invariant 
mass responses are not Gaussian, depend on 
photon conversions, and are different in different parts of the 
detectors. Therefore, the results shown below should be qualitatively valid 
but not quantitatively precise.} 
functions with mass resolution widths 
$\sigma_{\rm MR} = 1.3$, $1.5$, $1.7$, $2.0$ and $2.4$ GeV. 
This has the effect of reducing the peak and dip in the interference,
and moving their points of maximal deviations from 0 much farther from $M_H$.

\begin{figure}[!t]
\begin{minipage}[]{0.41\linewidth}
\begin{center}
\includegraphics[width=8.2cm,angle=0]{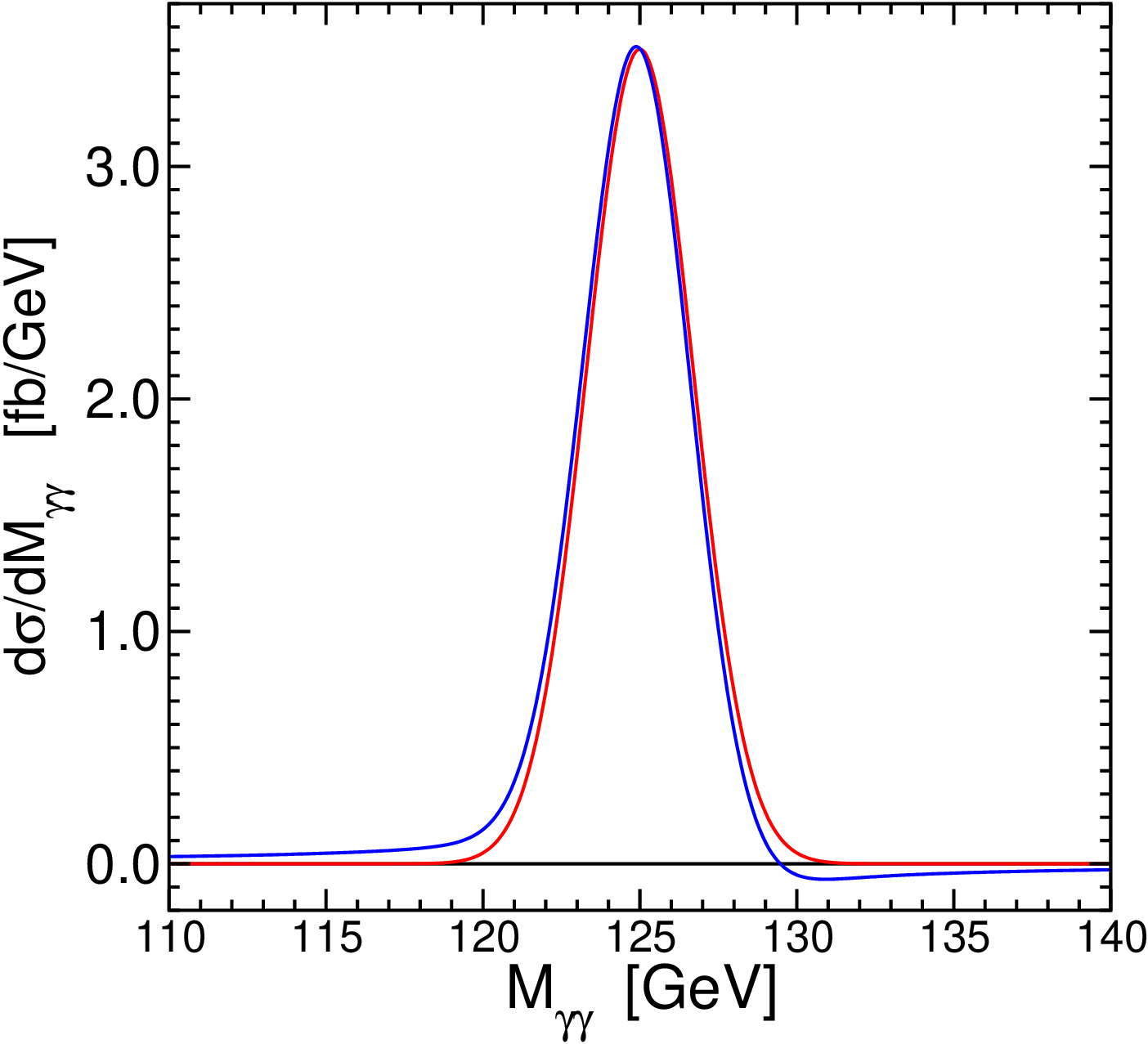}
\end{center}
\end{minipage}
\begin{minipage}[]{0.58\linewidth}
\begin{flushright}
\includegraphics[width=8.2cm,angle=0]{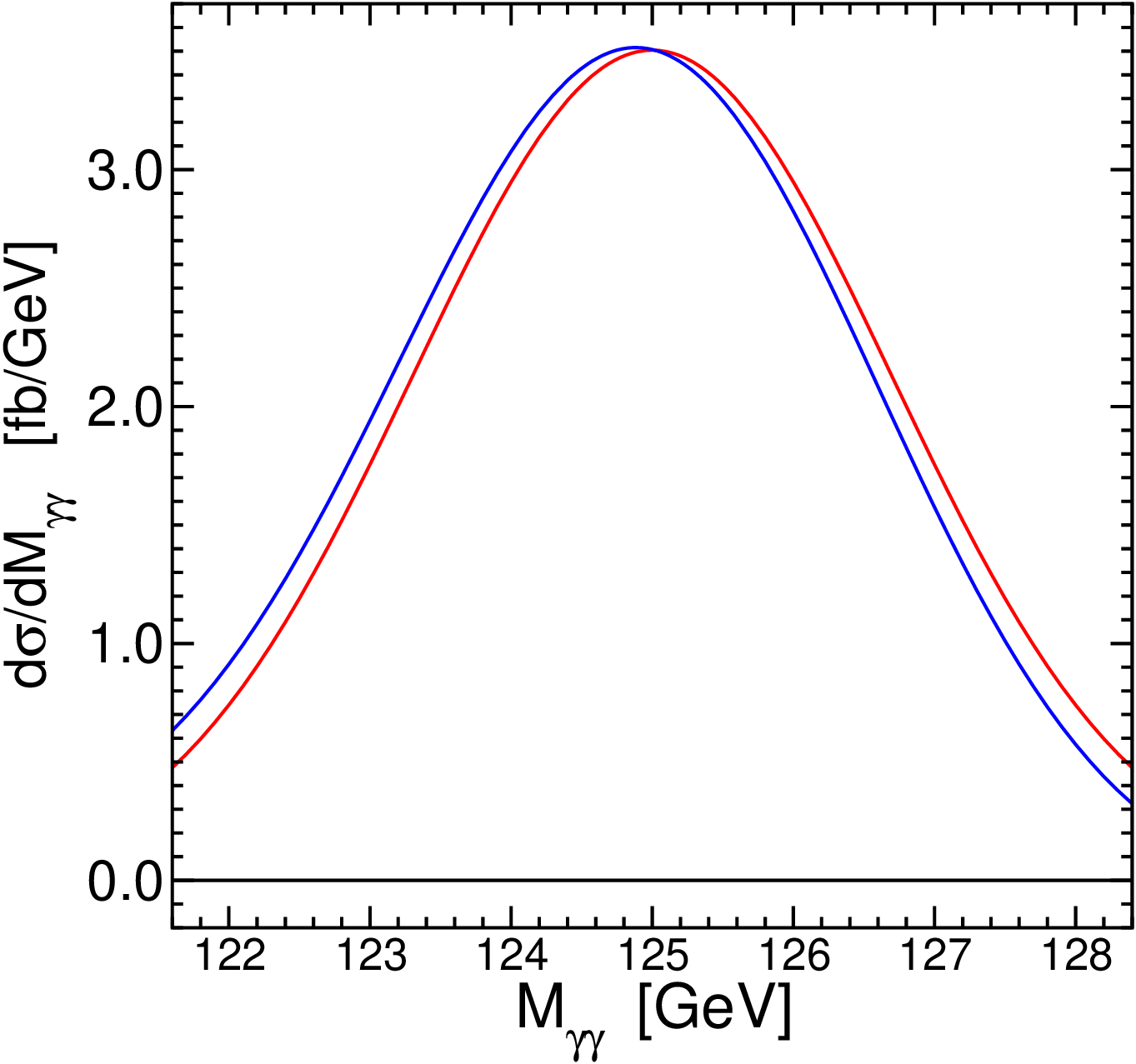}
\end{flushright}
\end{minipage}
\caption{\label{fig:smeared} Diphoton invariant mass distributions with a 
Gaussian mass resolution of width $\sigma_{\rm MR} = 1.7$ GeV. 
In each panel, the right (red) curve includes only the Higgs contribution 
without interference, and the left 
(blue) curve also includes the interference contribution from 
Figure \ref{fig:diffsmeared}. The right panel is a close-up of the left panel.} 
\end{figure}

To obtain the size of the shift in the Higgs peak diphoton
distribution, one can now combine the interference contribution with the
non-interference contribution from eqs.~(\ref{eq:sigmahint}) and (\ref{eq:NH}).
The results are shown in Figure \ref{fig:smeared} for the case of
a Gaussian mass resolution $\sigma_{\rm MR} = 1.7$ GeV. 
The distribution obtained including the interference effect is shifted slightly 
to the left of the distribution obtained neglecting the interference. 
In order to quantify the magnitude of the shift, it will be necessary to 
specify
the precise method used to fit the signal; this is again beyond the scope of the 
present paper. The background levels are
subject to significant higher order corrections
\cite{Binoth:1999qq,Bern:2002jx,Campbell:2011bn,Balazs:2006cc,Catani:2011qz}, 
and 
in practice are obtained 
by the experimental collaborations using a sideband analysis of 
fitting to the falling background
shape away from the Higgs peak. This fitting of the lineshape
to background plus signal will be affected by the slight surplus 
(deficit) of events below (above) $M_H$, depending on exactly how 
the fit is done.

\begin{figure}[!t]
\begin{minipage}[]{0.49\linewidth}
\begin{center}
\includegraphics[width=8.1cm,angle=0]{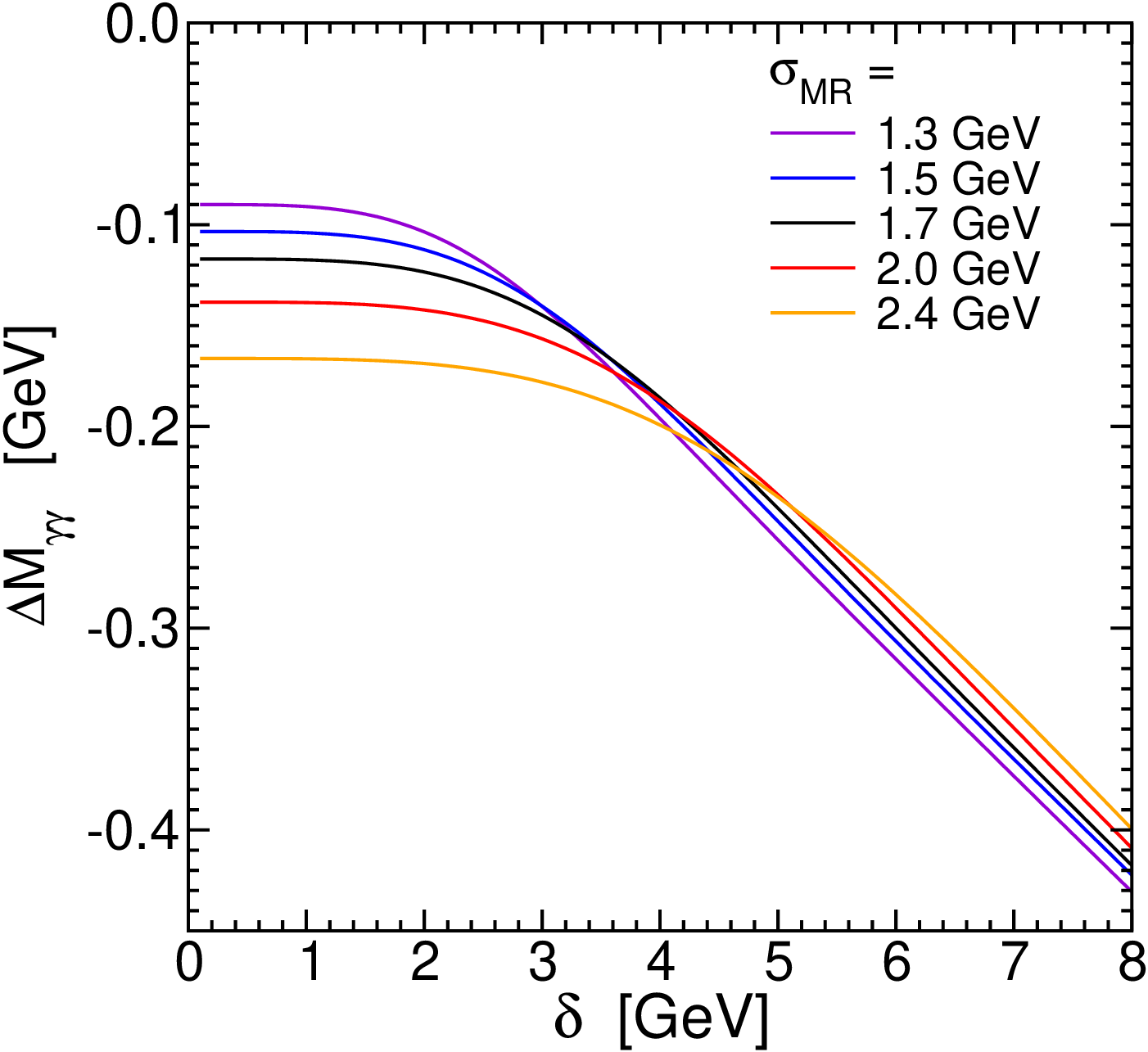}
\end{center}
\end{minipage}
\hspace{0.05\linewidth}
\begin{minipage}[]{0.44\linewidth}
\caption{\label{fig:deltam} The shift in the diphoton invariant mass 
distribution due to interference with the continuum background, 
using the measure of eqs.~(\ref{eq:defNdelta})-(\ref{eq:defDeltaM}),
for various assumed values of the mass resolution Gaussian 
width $\sigma_{\rm MR}$.}
\end{minipage}
\end{figure}

One simplistic way to estimate the shift is to take
a mass window $|M_{\gamma\gamma} - M_{\rm peak}| < \delta$, 
where $M_{\rm peak}$ is the invariant mass at the maximum of the 
distribution, and $\delta$ is supposed to be large enough to include 
most of the excess events over background in the peak, and then compute
\beq
N_\delta  &=& 
\int_{M_{\rm peak} - \delta}^{M_{\rm peak} + \delta} 
d M_{\gamma\gamma} \frac{d\sigma}{dM_{\gamma\gamma}},
\label{eq:defNdelta}
\\
\langle M_{\gamma\gamma} \rangle_{\delta} &=& 
\frac{1}{N_\delta} \int_{M_{\rm peak} - \delta}^{M_{\rm peak} + \delta} 
d M_{\gamma\gamma} \, M_{\gamma\gamma} \,\frac{d\sigma}{dM_{\gamma\gamma}} .
\label{eq:defMpeak}
\eeq
Now 
\beq
\Delta M_{\gamma\gamma} \equiv \langle 
M_{\gamma\gamma} \rangle_{\delta,\>\mbox{total}} 
- \langle M_{\gamma\gamma} \rangle_{\delta,\>\mbox{no interference}}
\label{eq:defDeltaM}
\eeq
is a theoretical measure of the shift due to including the interference.
For small $\delta$ ($\lsim 1$ GeV), $\Delta M_{\gamma\gamma}$ is essentially
just the shift in the maximum point of the distribution 
after subtracting background, which does not
correspond to an experimentally well-measured quantity. 
However, one can see 
from Figure \ref{fig:smeared} that including a 
wider window, which should be more similar to 
the methods used to determine $M_H$ by the experimental collaborations, will 
give a larger shift.
In fact, the magnitude of the shift $\Delta M_{\gamma\gamma}$ 
actually grows approximately 
linearly with $\delta$ for all $\delta \gsim 2\sigma_{\rm MR}$, 
due to the long positive (negative) tail at lower (higher) 
$M_{\gamma\gamma}$. This is shown in Figure \ref{fig:deltam},
where $\Delta M_{\gamma\gamma}$ is given as a function of $\delta$,
for various values of the Gaussian mass resolution $\sigma_{\rm MR}$.
Because a Gaussian mass resolution is assumed here for simplicity,
one finds
$\langle M_{\gamma\gamma} \rangle_{\delta,\>\mbox{no interference}} = M_H$
to very high precision, but 
$\langle M_{\gamma\gamma} \rangle_{\delta,\>\mbox{total}}$ is increasingly
smaller as $\delta$ is increased.
If one takes a value like $\delta = 4$ GeV as indicative, since this is large 
enough to include most of the signal events, 
then from 
Figure \ref{fig:deltam} the shift is about $-185$ MeV, with not much 
sensitivity to the assumed mass resolution. However, even a moderately
larger value of $\delta = 5$ GeV would increase the typical shift to about $-240$ MeV.

\begin{figure}[!t]
\begin{minipage}[]{0.49\linewidth}
\begin{center}
\includegraphics[width=7.3cm,angle=0]{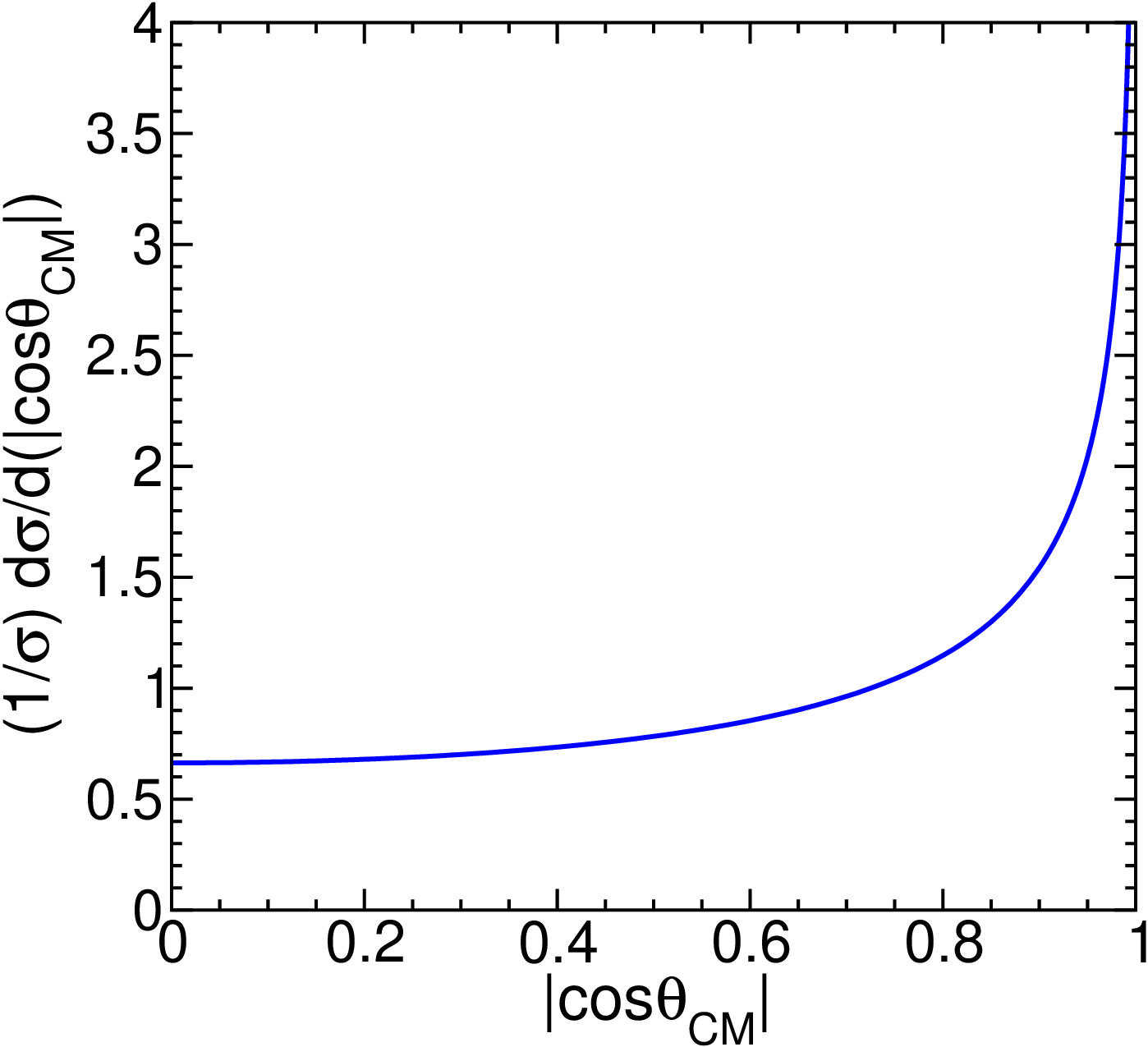}
\end{center}
\end{minipage}
\begin{minipage}[]{0.50\linewidth}
\begin{center}
\includegraphics[width=7.3cm,angle=0]{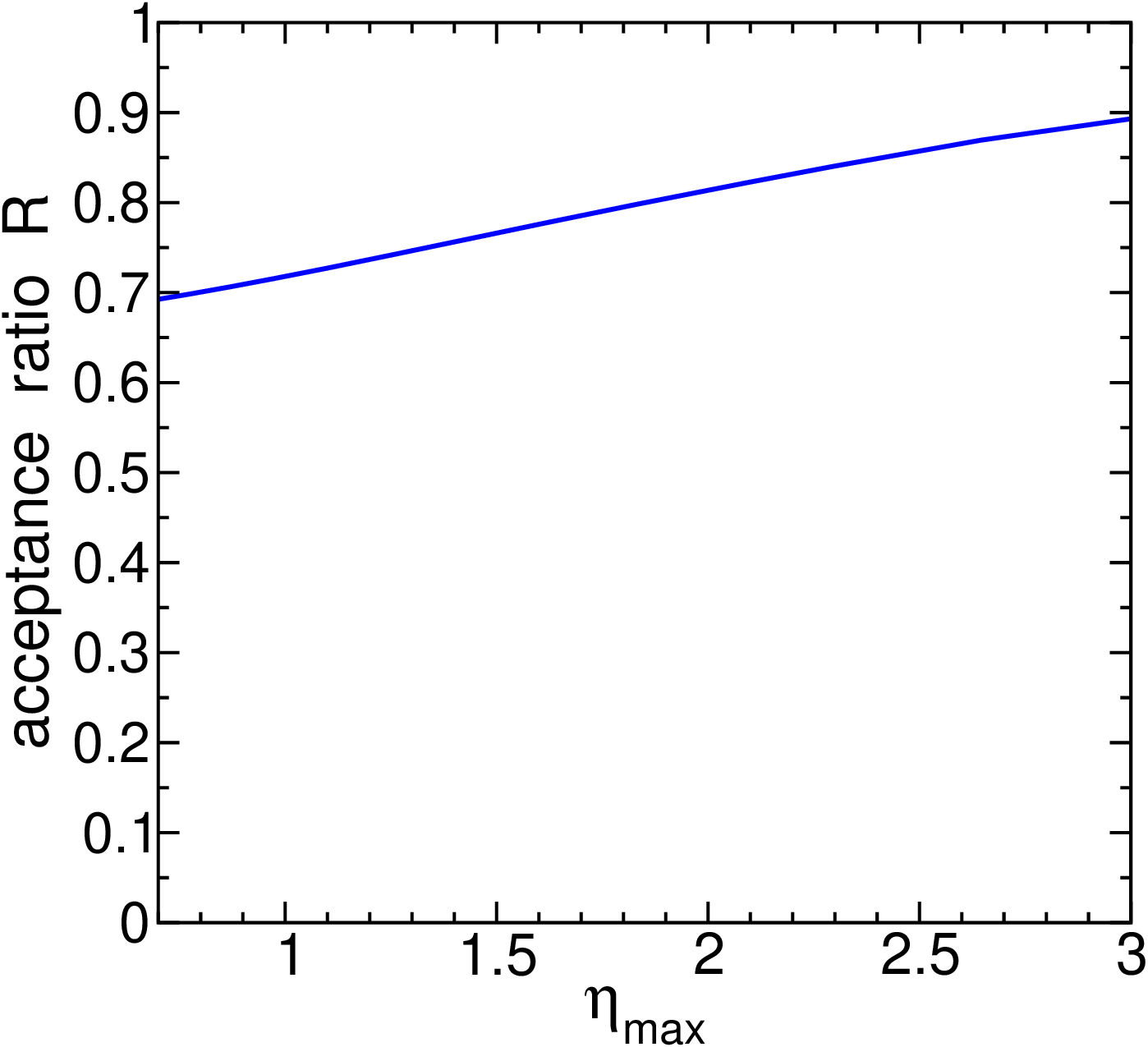}
\end{center}
\end{minipage}
\caption{\label{fig:angular} Angular distributions for the diphoton
Higgs signal-background interference. In the left panel, the shape
of the interference contribution
$(1/\sigma_{\rm int})d\sigma_{\rm int}/d(|\cos\theta_{\rm CM}|)$,
where $\theta_{\rm CM}$ is 
the diphoton center-of-mass scattering angle. In the right panel, the ratio 
of the acceptances
$R = (\sigma_{\rm cut}^{\rm int}/\sigma^{\rm int}_{\rm total})/
(\sigma^{H}_{\rm cut}/\sigma^{H}_{\rm total})$, where ``int"
refers to the Higgs-continuum interference part 
from eq.~(\ref{eq:NintR}) and ``$H$"
to the Higgs contribution without interference from 
eq.~(\ref{eq:NH}), and ``cut" means 
$|\eta| < \eta_{\rm max}$ for both photons, while ``total" means no cut on 
$\eta$.
} 
\end{figure}

\begin{figure}[!t]
\begin{minipage}[]{0.49\linewidth}
\begin{center}
\includegraphics[width=8.1cm,angle=0]{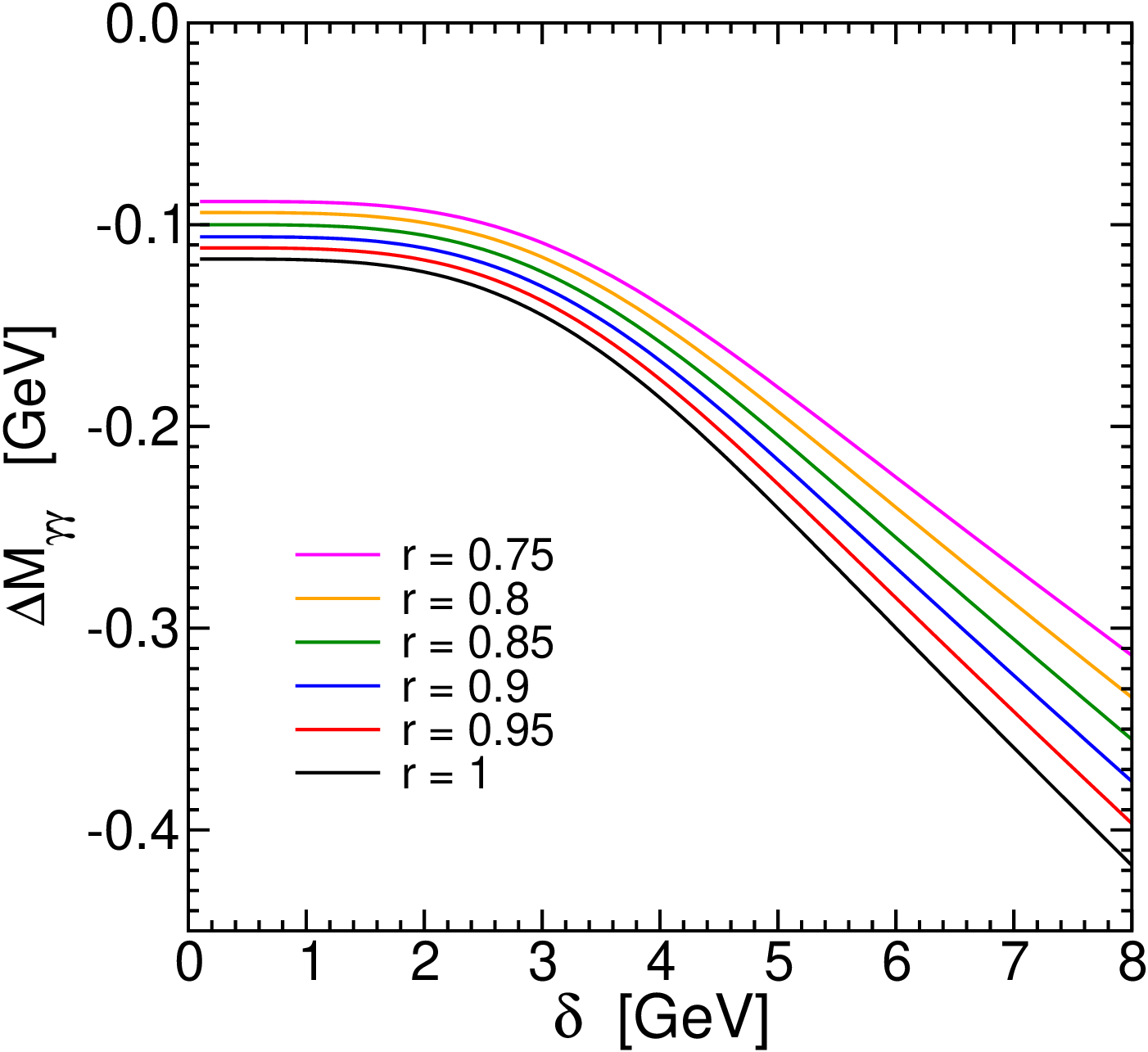}
\end{center}
\end{minipage}
\hspace{0.05\linewidth}
\begin{minipage}[]{0.44\linewidth}
\caption{\label{fig:deltamsupp} The shift in the diphoton invariant mass 
distribution due to the interference effect, 
using the measure of 
eqs.~(\ref{eq:defNdelta})-(\ref{eq:defDeltaM}) 
as in Figure \ref{fig:deltam}, but for a 
fixed mass
resolution $\sigma_{\rm MR} = 1.7$ GeV,
with the interference part of the total cross-section reduced by 
various factors $r$.}
\end{minipage}
\end{figure}

The results so far are based on total cross-sections, but experimental cuts
and efficiencies favor scattering into the central regions of the detectors.
In the CM frame, the non-interference part of the signal is isotropic, but the
interference is peaked at large $|z| = |\cos\theta_{\rm CM}|$,
as can be seen from eqs.~(\ref{eq:Mppmm}),
(\ref{eq:Mpppp}), (\ref{eq:NintR}), (\ref{eq:Aggaa}) and graphed in
the left panel of Figure \ref{fig:angular}. 
The way this angular distribution would 
translate into the effects of a cut on 
$\eta = -\ln[\tan(\theta_{\rm lab}/2)]$ is 
shown in the right panel of Figure \ref{fig:angular}. 
Here I show the ratio of 
acceptances
$R = (\sigma_{\rm cut}^{\rm int}/\sigma^{\rm int}_{\rm total})/
(\sigma^{H}_{\rm cut}/\sigma^{H}_{\rm total})$ 
as a function of $\eta_{\rm max}$, where ``int"
refers to the Higgs-continuum interference part from eq.~(\ref{eq:NintR})
and ``$H$"
to the Higgs contribution without interference from 
eq.~(\ref{eq:NH}), and ``cut" means 
$|\eta| < \eta_{\rm max}$ for both photons, while 
``total" means no cut on  $\eta$.
A simple cut on $\eta$ does not translate into experimental
reality, as the ATLAS Higgs analysis is sensitive to $|\eta| < 2.37$ except for
$1.37 < |\eta| < 1.52$, and CMS to $|\eta| < 2.5$ except for 
$1.44 < |\eta| < 1.57$, but with efficiencies that vary over those ranges. Both 
experiments also have cuts on the photon $p_T$'s, but the effect of this cannot 
be treated well by the present leading-order analysis. Furthermore, higher order 
corrections that have been neglected here could enhance or suppress the
interference part relative to the non-interference part. To illustrate the
possible effects of these considerations, Figure 
\ref{fig:deltamsupp} depicts the impact on the shift 
$\Delta M_{\gamma\gamma}$ of a relative suppression of the 
interference part of the cross-section 
by a factor of $r$. 
This shows that the effect of such a suppression
is to decrease the shift in the $M_{\gamma\gamma}$ peak by 
approximately the same factor $r$. For $r = 0.8$, 
the shift $\Delta M_{\gamma\gamma}$ found for $\delta = 4$ GeV 
would be reduced to about 150 MeV, although larger values are possible if
the signal-background 
fitting procedure effectively corresponds to larger $\delta$.

The measure of the mass shift used above is neither 
appropriate nor practical for use with real data, and does not correspond 
precisely to the techniques used by the experimental collaborations.  
However, the real lesson is that for a high precision determination 
of $M_H$, it will be necessary to fit to a signal lineshape that includes the 
interference effects. The leading-order 
estimates of this paper indicate that the
interference shifts the Higgs diphoton mass distribution lower by an amount of 
order 150 MeV compared to the expectation based on neglecting interference, 
depending on the method used to fit the data.
It would be useful to extend this analysis to include higher-order
contributions and realistic experimental cuts.
Although the shift is small, 
it is not negligible compared to the eventual precision that we may hope 
to obtain in the future, and to the last significant digit being
reported by the experimental collaborations for $M_H$ even now.
   
\bigskip \noindent 
{\it Acknowledgments:} I am grateful to the Aspen Center for Physics 
for hospitality and the National Science Foundation grant number 1066293.
This work was supported in part by the National Science Foundation grant 
number PHY-1068369.


\end{document}